\documentstyle{elsart}
\begin{document}
\begin{frontmatter}
\title{Shape-dependent universality in percolation}
\author[MI]{Robert M. Ziff\thanksref{EMAIL}}
\author[MI]{Christian D. Lorenz\thanksref{EMAIL2}}
\author[ME]{Peter Kleban\thanksref{EMAIL3}}
\address[MI]{Department of Chemical Engineering,
University of Michigan,
Ann Arbor, Michigan 48109-2136 USA}
\address[ME]{Laboratory for Surface Science and
Technology and Department of Physics and Astronomy,
University of Maine,
Orono, Maine 04469 USA}
\thanks[EMAIL]{email: rziff@umich.edu}
\thanks[EMAIL2]{email: cdl@umich.edu}
\thanks[EMAIL3]{email: kleban@maine.edu}

\begin{abstract}
The shape-dependent universality of the excess percolation cluster number
and cross-configuration probability on a torus is discussed.
Besides the aspect ratio of the torus, the universality class depends upon the twist in
the periodic boundary conditions, which for example are generally
introduced when triangular lattices are used in simulations.
\end{abstract}
\end{frontmatter}

\section{Introduction}

In standard random percolation, the universality of the critical
exponents is quite familiar.  The values of these universal quantities depend 
only upon the dimensionality of the system.  This is also    
the case for the amplitude ratios of quantities such as the
mean cluster size and correlation length that are defined essentially
for infinite systems.  Even though one might use finite systems
to measure such quantities, in the limit of the system length
scale $L \to \infty$, the shape of those finite systems becomes
irrelevant as $L$ exceeds the correlation length $\xi(p)$.

However, if a system is simulated exactly at $p_c$ where $\xi$ is 
in principle infinite,
or if one takes the scaling limit where $p$ approaches $p_c$ as $L$ is
increased such that $\xi$ remains a fixed fraction of $L$, then
the shape of the system (or the sequence of systems) becomes
relevant.  This leads to  another and larger
class of universal quantities, whose values depends upon the shape of the boundary
of the system and the boundary conditions, as well as the dimensionality.
  In spite of 
the resulting proliferation of classes, one still considers
these quantities to be universal, because they remain independent of the  ``microscopic"
features of the system (lattice type or continuum model used, whether
site or bond percolation, etc.).  

Examples of shape-dependent universal properties go back to the work of
Privman and Fisher \cite{PrivmanFisher} concerning Ising models on a torus. In percolation, the various
crossing problems considered by Langlands et al.\ \cite{Langlands92,Langlands94}, Aizenman \cite{Aizenman},
Cardy \cite{Cardy}, Ziff \cite{Ziff92}, Hovi and Aharony \cite{HoviAharony},
Hu et al.\ \cite{HuLin,HuLinChen}, Watts \cite{Watts} and Pinson \cite{Pinson}
show this type of universality.  These crossing quantites are all defined in terms of
macroscopic features of the system, so universality is clear.
Recently, Aharony and Stauffer \cite{AharonyStauffer}
have examined the shape dependence of critical universal ratios
such as $L^{-d}S / P_\infty^2$ where $S$ is the mean size (in sites)
of finite clusters and
$P_\infty$ is the probability a site belongs to the ``infinite"
cluster (one spanning the system, by some consistent definition).
Although thess quantities are defined on a microscopic (site) level,
the microscopic aspect cancels out by scaling theory
and the ratio is universal --- but again system-shape dependent.
Similar quantities for the Ising model
were studied by Kamieniarz and Bl\"ote \cite{KamieniarzBlote}
and M\"uller \cite{Muller}.

Here we discuss two shape-dependent universal
quantities: the excess cluster number and
the cross-configuration probability.  We 
also focus on twisted boundaries and the subtleties of the triangular lattice.

\section{The excess number of clusters}

Exactly at the critical threshold, the number of clusters per site or per unit area
 is a finite
non-universal constant $n_c$, whose values for site (S) and bond (B) percolation on square
(SQ) and triangular (TR) lattices were examined in detail in \cite{ZiffFinchAdamchik}. \
Here we quote those results in terms of the number per unit area, taking the lattice bond length to be
unity: 
$n_c^{S-SQ} = 0.027\,598\,1(3)$ and 
$n_c^{S-TR} = 0.020\,352\,2(6)$. \
For B-SQ, Temperley and Lieb \cite{TemperleyLieb} showed
\begin{equation}
n_c^{B-SQ} = \left[\left( -{\cot \mu \over 2} {\d \over \d \mu} \right)
\left\{{1\over 4\mu}\int_{-\infty}^\infty {\mathrm sech} \left( {\pi x \over2\mu} \right)
\ln \left( {\cosh x - \cos 2 x \over \cosh x - 1} \right) \d  x \right\}\right]_{\mu={\pi\over3}}
\end{equation}
which evaluates simply to \cite{ZiffFinchAdamchik}
\begin{equation}
n_c^{B-SQ} = {3 \sqrt 3 - 5 \over 2} = 0.098\,076\,211\ldots \ .
\end{equation}
Likewise,  Baxter, Temperley,  and Ashley's \cite{BaxterTemperleyAshley} integral
expression for $n_c^{B-TR}$ evaluates to  $(35/6 - 2/p_c^{B-TR})\sqrt 3$
$=0.129\,146\,645\ldots$ on a per-unit-area basis,
where $p_c^{B-TR} = 2\sin(\pi/18)$. \ Here,
clusters in bond percolation
are characterized by the number of wet sites they contain, with isolated sites corresponding to
clusters of size one.

While $n_c$ is non-universal, its finite-size correction is universal.
For an $L \times L'$ critical system with periodic boundary
conditions on all sides, the total number of
clusters $N$ behaves as 
$
N(L,L') =  n_c LL' + b + O( LL')^{-1}
$
where $b$ is a universal function of $r = L'/L$ \cite{ZiffFinchAdamchik}.  That is,
\begin{equation}
b(r) = \lim_{L \to \infty}[N(L,L') - n_c LL' ]
\label{eq:br}
\end{equation}
with $L' = rL$, and $b$ represents the excess number of
clusters over what one would expect from the bulk density.  As such, it
 reflects the large clusters of the system, which is the basis of
its universality.  Aharony and Stauffer \cite{AharonyStauffer} showed
that the universality of $b$ follows directly from the arguments of Privman and Fisher \cite{PrivmanFisher}
applied to percolation, implying that $b$ is
precisely the value of the free energy scaling function at $p_c$ and $h=0$. \  Note
that this ``free energy" is  $\sum_s n_s e^{-hs}$, different from the free energy
given below.

In  \cite{ZiffFinchAdamchik}, $b(1)$ was numerically found to equal 0.884 for both site and bond
percolation  on a SQ lattice, demonstrating universality.
It was also found that $b(2) = 0.991$, $b(4) = 1.512$, and for large $r$ (systems of very high aspect ratio),
 $b(r) \sim \tilde b \, r$ 
with $\tilde b = 0.3608$. Here  $\tilde b$ is
the excess number per unit length along an infinite periodic strip or
cylinder.  Periodic b.\ c.\ are essential in this
problem to eliminate boundary effects which would otherwise overwhelm $b$.

In \cite{KlebanZiff} it was shown that $b$ can be found
explicitly from exact coulomb-gas results.  In the Fortuin-Kastelyn
representation, the partition function of the Potts model at criticality
is $Z = \sum Q^{N_C+N_B/2}$ where $N_C$ is the number of clusters
and $N_B$ the number of bonds, giving bond percolation
in the limit of $Q = 1$.  Here the free energy is 
$F = \ln Z$, and $\langle N_c + N_B/2 \rangle$ follows from  $dF/dQ$ at
$Q=1$.  Using the Potts model
partition function (universal part) of Di Francesco et
al.\ \cite{DiFrancescoSaleurZuber},
\cite{KlebanZiff} obtained  
\begin{eqnarray}
b(r) & =  { 5\sqrt{3}\,r \over 24} + q^{5/4}(2\sqrt3\,r-\half) 
+q^2(\sqrt3\,r-1) 
+ q^{5/48} + 2q^{53/48}\nonumber \\ 
& -q^{23/16}+q^{77/48}+\ldots
\label{eq:b(r)}
\end{eqnarray}
where $q=\e^{-2 \pi r}$. \
This result yields $b(1) = 0.883\,576\,308\ldots$, $b(2) = 0.991\,781\,515\ldots $
$b(4)  = 1.516\,324\,734\ldots $, and $\tilde b = 5\sqrt3/24$,
consistent with measurements of \cite{ZiffFinchAdamchik}.
The result for $\tilde b$ also follows
directly from the work of Bl\"ote, Cardy and Nightingale
\cite{BloteCardyNightingale} on cylindrical systems.
Fluctuations and higher-order cumulants
are also discussed in \cite{KlebanZiff}, and $\tilde b$
in 3d is discussed in \cite{LorenzZiff}.

\section{Confirmation of universality using triangular lattices}

Demonstrating universality of $b(r)$ by choosing only B-SQ and S-SQ
as in \cite{ZiffFinchAdamchik} may not be completely convincing,
and one would like to compare, say, a TR and SQ lattice.  Periodic
b.\ c.\ must be retained.  Two obvious ways to represent the TR
lattice on the square array of a computer program are shown in Fig\ 1. \
In (a), the TR lattice topology is created by choosing
every other site of the SQ lattice.  Taking an $L \times 2L$ boundary on the SQ
lattice as shown, the effective boundary on the TR lattice becomes a rectangle
with $r=\sqrt3/2$. \  Periodic b.\ c.\  on the underlying
SQ lattice results in normal (untwisted)
periodic b.\ c.\  on the TR lattice also. 
The measured $b = 0.887$ for this system agrees completely
with the theoretical $b(\sqrt3/2) = 0.887\,373\,266$
for a rectangular system with aspect ratio $\sqrt{3}/2$.

\begin{figure}
\vspace{70mm}
\caption{Two representations of a triangular lattice on
a square array, yielding (a) a rectangular boundary of
aspect ratio $\sqrt{3}/2$ with
no twist, and (b) the same rectangular boundary but with a twist 
of $1/2$.}
\end{figure}

The second obvious way to represent the TR lattice --- by far the most
common one --- is shown in Fig.\ 1(b). \  Diagonals are
simply added to the SQ lattice, and the periodic
b.\ c.\  are applied to the squared-off lattice as is.  Making this into a proper
TR lattice, the system becomes a $1 \times 1\ 60^\circ$ rhombus, and shifting
around the triangle as shown in Fig.\ 1b demonstrates that it is effectively 
a rectangular boundary with $r=\sqrt3/2$, but with a ``twist" $t=\half$ in the periodic
b.\ c., meaning that the $x$ coordinates are shifted by a fraction $t$ of the total
length when wrapping around in the vertical direction.  For this
system, simulations gave $b(\sqrt3/2,1/2) = 0.878$  \cite{ZiffFinchAdamchik}, less
than $b(\sqrt3/2,0) = 0.8874$ and indeed less than the minimum untwisted rectangle,
$b(1,0) = 0.8836$, where now we write $b = b(r,t)$.

Here we demonstrate the universality of $b(\sqrt3/2,1/2)$
by studying a SQ lattice with $r$ (necessarily rational) close to $\sqrt3/2 \approx 0.866$,
and comparing the results to the above TR lattice measurement.
On the SQ lattice, we considered a system of size $14\times 16$,
where $r=0.875$; the measured values of $b$ for $t=0,\ 1/8,\ldots,\ 1$ are shown in Fig.\ 2. \
At twist $\half$, the value of $b$ is very close to the result 0.878 found on the TR lattice,
\cite{ZiffFinchAdamchik} demonstrating the universality between these two lattices.
Note that, to find $b(\sqrt 3/2,1/2)$ on the SQ lattice to high precision, one would have to consider different
size systems and extrapolate to $\infty$, and different rational $r$ to interpolate to $r = \sqrt3/2$.
 Results for a square system
$16\times16$ are also shown in Fig.\ 2.

\begin{figure}
\vspace{70mm}
\caption{Excess cluster number $b(r,t)$ vs. twist $t$ for
$14 \times 16$ simulation (triangles),
theory for $r = \sqrt{3}/2$ (solid line),
$16 \times 16$ simulation (squares) and
theory for $r = 1$ (broken line).}

\end{figure}

We have generalized the theoretical methods described in \cite{KlebanZiff} to find
$b(r,t)$ from the partition function of \cite{DiFrancescoSaleurZuber}.  The parameter
$\tau$ becomes $t + i r$, and the results, which are rather involved, yield the
solid curves in Fig.\ 2. \ The discrepancy with the numerical values can
be attributed to the small system size of these simulations.

\section{Symmetries on a torus with a twist}

The torus with a twist has various topological symmetries
that apply to any shape-dependent universal quantity $u(r,t)$,
which includes $b(r,t)$.  We consider a rectangular boundary
with base 1 and height $r$, with a horizontal twist $t$ in the
periodic b.~c. \   (Note that having twists in two directions leads to a 
non-uniform system, so we don't consider it.) \   $u(r,t)$ satisfies the obvious symmetries
of reflection
\begin{equation}
u(r,t) = u(r,-t)
\label{eq:reflection}
\end{equation}
and  periodicity in the $t$ direction
\begin{equation}
u(r,t) = u(r,1+t)
\label{eq:periodicity}
\end{equation}
Another symmetry follows from the
observation that the same rhombus can be made into a rectangle
in two different ways, leading to:
\begin{equation}
u(r,t) = u\left({r \over r^2 + t^2},{t \over r^2 + t^2}\right)
\label{eq:inverse}
\end{equation}
Another construction
shows that when  $t = 1/n$ where $n$ is an integer,
\begin{equation}
u\left(r,{1 \over n}\right) = u\left({1\over n^2 r},{1 \over n}\right)
\label{eq:integer}
\end{equation}
which also 
follows from
Eqs.\ (\ref{eq:reflection}-\ref{eq:inverse}).
On the complex $\tau = t + ir$ plane, (\ref{eq:inverse}) corresponds to $\tau \to 1/\tau$ while
 (\ref{eq:periodicity}) corresponds to $\tau \to \tau + 1$.  These transformations
generate the modular group, and functions invariant under them 
are called modular.  Thus, $b(r,t)$ must necessarily be a modular function.
However, the explicit expression for $b$ does not display that modularity clearly.

Besides the excess number, another universal
 quantity on a torus is the cross-configuration probability
$\pi_+(r,t)$, which can be expressed in a quite compact form.
Using the results of \cite{DiFrancescoSaleurZuber}, 
Pinson \cite{Pinson} has shown $\pi_+(r,t) =
\half [Z_c(8/3)-Z_c(2/3)]$, where 
\begin{equation}
Z_c(h) = {\sqrt{h/r}\over
\eta(q)\overline\eta(q)}\sum_{n,n'}
\exp\left\{-{\pi h\over r}[n'^2+n^2(r^2+t^2)-2tnn']\right\} \ ,
\label{eq:pinson}
\end{equation}
$\eta(q)$ is the Dedekind eta function and $q = e^{-2\pi (r - it)}$. 
It can be easily verified
that this function satisfies the modular symmetries.
For an untwisted torus (\ref{eq:pinson}) reduces to
\begin{equation}
 \pi_+(r,0) = {1 \over 4} \sqrt{3 \over 2} \ 
{\varphi({3r\over 8}) \varphi({3 \over 8 r}) 
- 2\varphi({3r\over 2}) \varphi({3 \over 2 r}) \over \tilde\eta(r)
\tilde\eta({1\over r}) } 
\label{eq:pir}
\end{equation}
where $\tilde\eta(r)=\eta(e^{-2\pi r})$ and $\varphi(r) = \vartheta_3(e^{-\pi r})$
is the Jacobi theta function in Ramanujan's notation.
The symmetry $r \to 1/r$ is apparent.
For all rational
$r$, $\pi_+(r,0)$ is an algebraic number; for example, for $r=1$, 
one can show
\begin{equation}
\pi_+(1,0) = {2 \sqrt{a + \overline b} \sqrt{2\sqrt{a\overline b}}
- (a + \overline b) + 2\sqrt{a\overline b} \over   3^{1/4}\ 4}
= 0.309\,526\,275\ldots
\end{equation}
where $a = 1 + \sqrt 3 $ and $\overline b = \sqrt 2 - 3^{1/4}$,
using various results for theta functions. 
According to the symmetries above, this same value $\pi_+(1,0) = 0.3095\ldots$ 
applies to 
$(r,t) = (1/2,1/2) = (1/5,3/5) = (1/10,3/10)
= (1/13,5/13) = (1/17,13/17)  = (1/26,5/26) =
(1/29,12/29) = (1/34,13/34) = (1/37,6/37) = (1/41,9/41) = (1/50,14/50)$
and infinitely many other systems, just as 
$b(1,0) = 0.8836\ldots$ applies to all these systems.

We have also measured $\pi_+$ in the $14\times16$ and $16\times16$ systems, with various $t$. \
Whether crossing occurs can be found using an indicator function, such as $I = N_C - N_{C'}
+ N_B - N_S$, where $N_C$ is the number of clusters, $N_{C'}$ is the number of dual-lattice
clusters, $N_B$ is the number of bonds, and $N_S$ is the number of sites.  When $I=1$, there is
a cross-configuration on the lattice, when $I=-1$ there is a cross configuration
on the dual-lattice (these two events are clearly mutually exclusive), and when $I=0$ there
is neither.  In the latter case, there will necessarily be at least one wrap-around cluster
on the lattice or dual-lattice, or a spiral.  Another indicator function
can be made using the number of hulls in the system \cite{Pinson}.

In Fig.\ 3 we show the measured $\pi_+(r,t)$ and comparison with
predictions of Pinson's formula.  The small deviations are presumably due to finite-size
effects which should disappear when larger systems are measured and an extrapolation
is made to infinity, as we have verified for $t=0$.  Note that $\pi_+(\sqrt{3}/2,1/2) 
= 0.316\,053\,413\ldots$ is at a local and apparently global maximum of Pinson's formula,
although again  by equations (\ref{eq:reflection}-\ref{eq:inverse})
an infinite number of other points on the $(r,t)$ plane have the
identical value of $\pi_+$, such as $(\sqrt3/6,1/2)$ and  $(\sqrt3/14,5/14)$.

\begin{figure}
\vspace{70mm}
\caption{Pinson's number $\pi_+(r,t)$ vs.\ twist $t$; legend same
as Fig.\ 2.}
\end{figure}

A plot of Pinson's formula as a function of $t$ for different
$r$ shows that for small $r$ it begins to develop oscillations.  This is
because of the tendency to create spiral rather than cross
configurations for small aspect ratios.

\section{The meaning of $b$}

In \cite{ZiffFinchAdamchik}, it was suggested that $b$ relates to the
number of ``spanning" clusters,
since these are essentially the cause of the excess.
However, Hu \cite{Hu} has shown that they are not numerically identical, using
one particular definition of spanning.  Here we elaborate on this point.

We consider large systems 
with periodic b.\ c.\ and ask for the number of clusters Nr$(\ell_m \ge \ell)$
whose maximum
dimension $\ell_m$ in the $x$ or $y$ direction exceeds some value $\ell$. \
Since $s \sim \ell_m ^D$ where $D$ is the fractal dimension and $n_s \sim s^{-\tau}$,
it follows that  Nr$(\ell_m \ge \ell) \sim \ell^{(1-\tau)D} = \ell^{-d}$
or $\ell^{-2}$ in 2d. \ We have measured this quantity for square $L\times L$  and rectangular $L\times
2L$ systems for various sizes, and find that for an intermediate range in $\ell$ the expected universal
$\ell^{-2}$ behavior is followed:
\begin{equation}
{\mathrm Nr}(\ell_m \ge \ell) = C \left( {\ell \over \sqrt{A} } \right)^{-2}  = CA/\ell^2
\label{eq:Nr}
\end{equation}
where $A = L^2$ (square) or $2L^2$ (rectangular system) and $C=0.116$. \ $C$ is a universal measure
of the size distribution, dependent only upon the rule of what constitutes the length scale $\ell$. 
(One could just as well use maximum diameter, radius of gyration, etc., each of which would lead to a
different
$C$.  Thus $C$ too is a ``shape"-dependent universal quantity.)   Eq.~(\ref{eq:Nr}) implies that $C$ is
just the density in an infinite system of clusters of minimum dimension $\ell$, on the length scale of
$\ell$. It is the universal analog of $n_c$.

The data for Nr$(\ell_m \ge \ell)$ deviates
 from Eq.\ (\ref{eq:Nr}) at both the large and
small size limits. For small $\ell$, the deviation is due to lattice-level effects;
at the limit $\ell = 1$, Nr$(\ell_m \ge 1)$ is just
$n_c A$, which is clearly non-universal.  For $\ell$ near the maximum, the deviation is due to the influence of the boundary
and is related to the value of $b$.

According
to our definition, $b = $ (the actual number of clusters) $-$ (the expected number of clusters using
the bulk density).  Therefore, using a lower length scale of $\ell$
that is in the scaling region, we have
\begin{equation}
b = {\mathrm Nr}(\ell_m \ge \ell) - C \, A/\ell^2
\label{eq:b3}
\end{equation}
Evidently, in using this formula, $\ell$ can be taken right up to the minimum dimension of the system, $L$.
We find that the number in each size range,
$N_\ell = {\mathrm Nr}(\ell/2 \le \ell_m < \ell)$ $ = {\mathrm Nr}(\ell_m \ge \ell/2)  - {\mathrm Nr}(\ell_m \ge \ell)$
follows $N_{\ell/2}/N_{\ell}  = 4$ within $\pm 0.01$ right up to $\ell = L$. \
 This implies that 
$b$ can be found by applying (\ref{eq:b3})
with $\ell = L$.  \
Using the numerical data for Nr$(\ell_m \ge L)$ for $r=1$ and $r=2$,
we find from (\ref{eq:b3})
\begin{eqnarray}
& b(1,0) & = 0.990 - 0.116 = 0.87 \nonumber \\
& b(2,0) & = 1.214 - 2(0.116) = 0.98 
\label{eq:num}
\end{eqnarray}
compared with the actual vales 0.884 and 0.991 respectively.  The small shortfall may be due to the
need to take $\ell$ somewhat smaller than $L$ in (\ref{eq:b3}), or 
to statistical errors.  We are investigating this point further.

Thus, we have the meaning of $b$ (taking $\ell = L)$: the number of clusters in a system of area $A$ 
whose extent is larger than  the minimum system
dimension
$L$, minus the expected number predicted by the bulk density,  $C A/L^2$.

\section{Conclusions}

We have numerically demonstrated the universality of $b(\sqrt3/2,1/2)$
on both the SQ and TR lattices.  We have shown that $b$ is related
to the average number of clusters of length scale greater or equal to $L$, if one
subtracts off the contribution of the universal size distribution
characterized by the universal constant $C$.

It appears that $N_\ell$ follows a universal behavior close to, and perhaps right
up to $\ell = L$. On a finite system with periodic b.\ c., the
probability of growing a cluster of a certain size is identical to its probability
on an infinite system, as long as the cluster is small enough that it doesn't
touch itself after wrapping around the boundaries.  This result, however, says that
the {\it number} of clusters of a certain
size range is also substantially unaffected by the system finiteness,
even though the boundaries should, it seems, influence
the statistics of such clusters when their combined size is large
enough that they touch when wrapping around the
boundary.  Further work needs to be done to understand this behavior.

The TR lattice constructed as in Fig.\ 1b gives extremal values of $b$ and $\pi_+$
and may in fact be the best periodic system to use for many percolation problems (as well as other
types of lattice simulations).  This is because, when the $1 \times 1$ $60^\circ$ rhombus is
 used to tile the plane, it leads to a triangular array of repeated patterns which has the
most space between each repeated element of any regular array.

While tori with ``twists" add a rich extra degree of freedom,
they are by no means the only systems for which $b$ can be calculated.
What is needed is a system that is effectively a closed surface.
One can transform the rectangular basis of the torus to other shapes
by conformal transformation, such as to an annulus,
and then apply the transformed periodic b.\ c. to the problem.
(Here, the curved boundaries
suggest using a continuum form of percolation, as in \cite{HsuHuangLing}).
A simple closed surface like the surface of a
sphere can also be used for the system.  Each of these systems will have its own characteristic
value of $b$ and other shape-dependent universal quantities.

\ack
RZ acknowledges NSF grant
DMR-9520700. Correspondence with  K. S. Williams and B. Berndt concerning
theta function identities is gratefully acknowledged.

\vfill
\eject

\begin{thebibliography}{9}


\bibitem{PrivmanFisher} V. Privman and M. E. Fisher, Phys. Rev. B {\bf 30}, 322
(1984).

\bibitem {Langlands92} R. P. Langlands, C. Pichet, P. Pouliot, and Y. Saint-Aubin, J. Stat. Phys. {\bf 67}, 553 (1992).

\bibitem {Langlands94} R. P. Langlands, P. Pouiliot, and Y. Saint-Aubin, Bull.
AMS {\bf 30}, 1 (1994).

\bibitem{Aizenman} M. Aizenman, in The IMA Volumes in Mathematics and
its Applications (Springer-Verlag, 1997).

\bibitem {Cardy} J. L. Cardy, J. Phys. A: Math Gen. {\bf 25}, L201 (1992).

\bibitem{Ziff92} R. M. Ziff, Phys. Rev. Lett.{\bf 69} 2670
(1992).

\bibitem {HoviAharony} J.-P. Hovi and A. Aharony, Phys. Rev. E {\bf 53}, 235 (1996).

\bibitem {HuLin} C.-K. Hu and C.-Y. Lin, Phys. Rev. Lett. {\bf 75}, 193 (1995). 

\bibitem {HuLinChen} C.-K. Hu, C.-Y. Lin and J.-A. Chen, Phys. Rev. Lett. {\bf 77}, 8
(1996).

\bibitem {Watts} G. M. T. Watts, J. Phys. A: Math Gen. {\bf 29}, L363 (1996).

\bibitem{Pinson} H. T. Pinson, J. Stat. Phys. {\bf 75}, 1167 (1994).

\bibitem{AharonyStauffer} A. Aharony and D. Stauffer, J. Phys. A: Math Gen. {\bf 30}, L301 (1997).

\bibitem {KamieniarzBlote} G. Kamieniarz and H. W. J. Bl\"ote, J. Phys. A
{\bf 26}, 201 (1993).

\bibitem {Muller} B. M\"uller, Int. J. Mod. Phys. C {\bf 9}, 1 (1998). 

\bibitem{ZiffFinchAdamchik} R. M. Ziff, S. R. Finch, and V. S. Adamchik, Phys. Rev.
Lett., {\bf 79} 3447 (1997).

\bibitem{TemperleyLieb} H. N. V. Temperley and E. H. Lieb,
Proc. R. Soc. Lond. A. {\bf 322}, 251 (1971).

\bibitem{BaxterTemperleyAshley} R. J. Baxter, H. N. V. Temperley, and S. E. Ashley,
Proc. R. Soc. Lond. A. {\bf 358}, 535 (1978).

\bibitem{KlebanZiff} P. Kleban and R. M. Ziff, Phys Rev. B {\bf 57}, R8075 (1998).

\bibitem{DiFrancescoSaleurZuber} P. Di Francesco, H. Saleur, and J. Zuber,
J. Stat. Phys. {\bf 49}, 57 (1987).

\bibitem{BloteCardyNightingale} H. Bl\"ote, J. Cardy, and M. Nightingale,
 Phys. Rev. Lett.{\bf 56} 742 (1986).

\bibitem{LorenzZiff} C. D. Lorenz and R.M. Ziff, Phys. Rev. E {\bf 57}, 230
(1998).

\bibitem {Hu} C.-K. Hu, Rutgers Statistical Physics Meeting, December 1997. 


\bibitem {HsuHuangLing} H.-P. Hsu, M.-C. Huang, and K.-J. Ling, Phys. Rev. B {\bf 56}, 10743 (1997). 


\end{thebibliography}
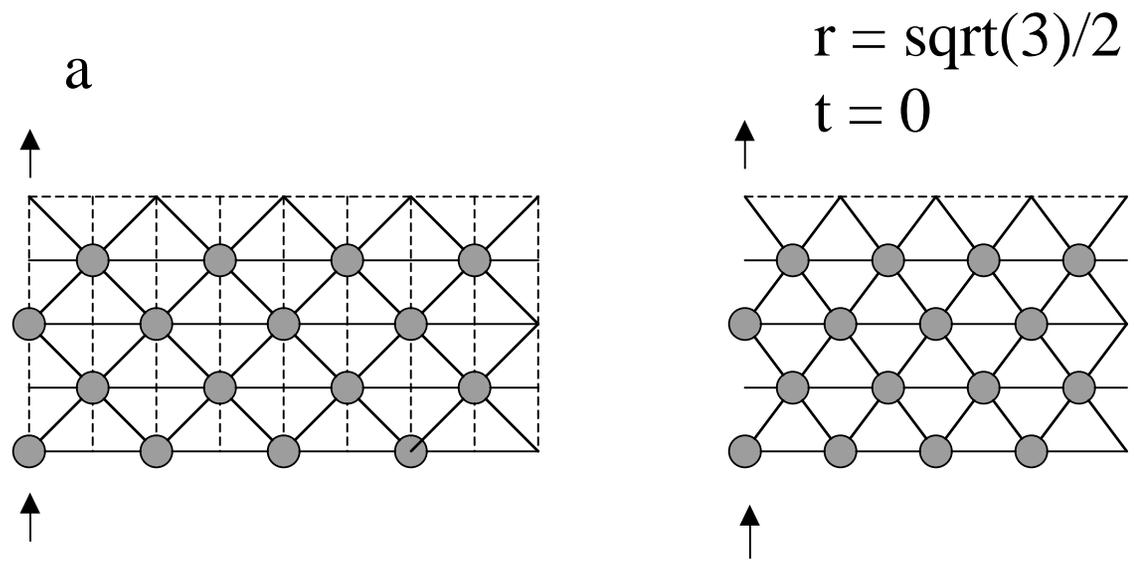
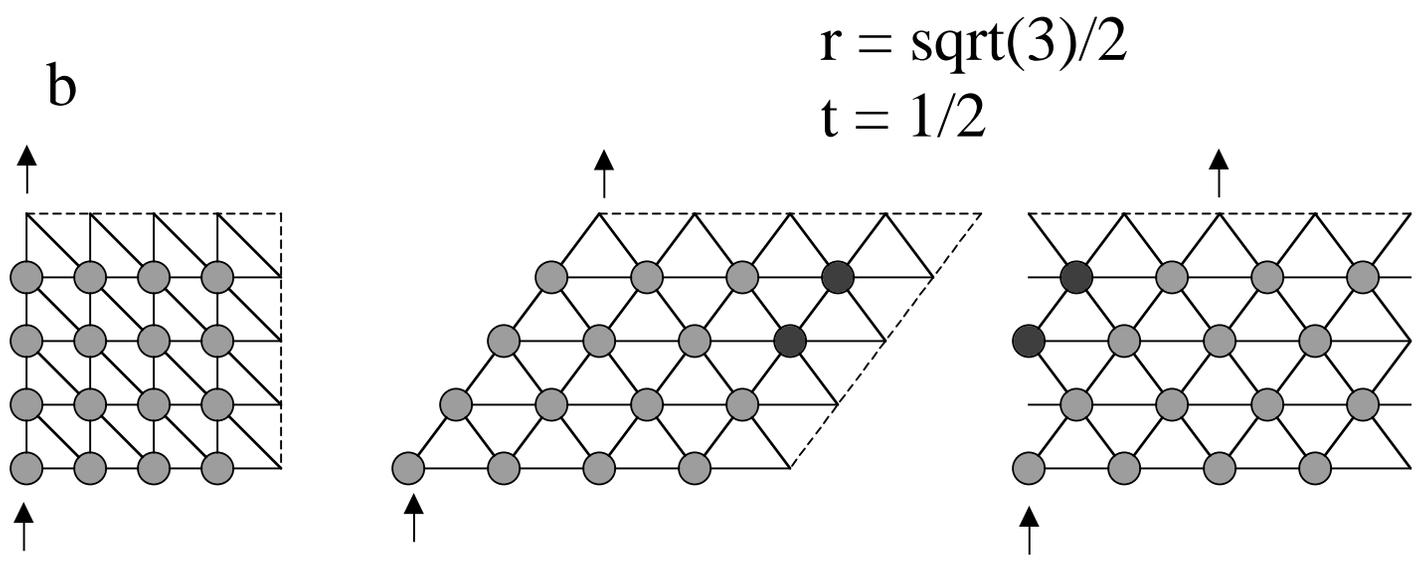
\end{document}